\documentstyle[12pt,epsfig]{article}
\begin{document}

\def\thefootnote{\fnsymbol{footnote}}

\begin{flushright}
PM--97--04\\
FSTT--97-03\\
hep-th/9703xxx\\
\end{flushright}

\vspace{1cm}

\begin{center}

{\large\sc {\bf Note on the deformation of chiral algebra. }}

\vspace{1cm}

{\sc A. Arhrib$^{1,2}$\footnote{Permanent address: D\'epartement de math\'ematiques 
Facult\'e des Sciences et Techniques , B.P 416 Tanger, Morocco}}, 
{\sc M. Rachidi$^{3}$ } and  {\sc E. H. Saidi$^{4}$ }
 
\vspace{1cm}

$^1$ Laboratoire de Physique Math\'ematique et Th\'eorique, UPRES--A 5032,\\
Universit\'e de Montpellier II, F--34095 Montpellier Cedex 5, France.

\vspace*{3mm}

$^2$ Laboratoire de Physique et Techniques  Nucl\'eaires\\   
Facult\'e des Sciences Semlalia, B. P S15, Marrakech Morocco.

\vspace*{3mm}

$^3$ D\'epartement de Math\'ematiques, Facult\'e des Sciences \\ 
Av Ibn Battouta, B.P 1014 
Rabat, Morocco 

\vspace*{3mm}

$^4$ Section de Physique des Hautes Energies, LMPHE, 
Facult\'e des Sciences,\\ Av Ibn Battouta, B.P 1014 
Rabat, Morocco

\end{center}

\vspace{2cm}

\begin{abstract}
We introduce a new type of deformation of the chiral symmetry
based on the deformation of the Laurent expansion of the conformal
energy momentum tensor. Two kinds of solutions of 
the deformed equations of continuity are worked out. 
Known results are recovered, others features are also discussed.
\end{abstract}

\newpage
 
\section{Introduction}
In the last decade, a great attention has been given to the  study of
two dimensional
conformal  invariance and its extensions \cite{polyakov,zamolod,fateev}.
These symmetries play  a central
role in the string theory, two dimentional gravity and
 statistical models of critical phenomena.
 
Recently some interest has been devoted to the deformation of
these infinite dimentional symmetries. This is motivated essentially
by    the fact that
some deformed symmetries are believed to be hidden symmetries
responsible   of the
integrability of many massives $2d$ models \cite{mussorado}.  Before
presenting our problem, let us
first recall our way of understanding what is meant by the idea of
deformation of
symmetries. Roughly speaking there are various ways to deform
symmetries which basically can be classified in two kinds.
\begin{itemize}
\item The q--deformation \cite{kadiri, leclari,chaichian} which consists
to deform the Lie algebra structure $l(s)$ of the usual bracket
\begin{equation}
[A,B]=AB-BA\ \ \  as \ \ \ \  [A,B]= AB-q BA
\end{equation}
In this eq. $A$ and $B$ are two elements of a Lie algebra $g$ and $q$
is a generic
complex parameter. For $q=1$, one get the unperturbed Lie bracket and
for the special
case $q=-1$, one obtains the anticommutator of supersymmetric algebra.
In the general
case the generic deformed bracket (1) is valued in $U_q(g)$, the
universal Lie algebra of $g$. Generalizations of this deformation have been also studied
\item The conformal deformation which deals with the deformation of
the   conformal
(complex) structure $c(s)$ of two dimensional scale invariant theories.
These deformations are governed by the Zamolodchikov c--theorem   for
Crossover phenomena \cite{zamolodc}. Massive deformations are expected
to  be governed by some generalization of this theorem.
\end{itemize}
 
Note that the
conformal and Lie algebra deformations are two independent operations,
that is:
\begin{equation}
c(s) l(s)=l(s) c(s)
\end{equation}
Note also that both the deformation of the Lie algebra structure eq(1)
and the
conformal one were shown to be powerful ideas to connect different
structures which
were viewed before as independent. The    connection of graded Lie
algebras   to the
standard bosonic ones \cite{zhang} and the connection of the
integrability structure to the
conformal one \cite{mussorado} are may be two good lecons that we have
been thought from the two types of deformations mentioned above.
 
In this paper we want to make some comments on the conformal deformation.
More
precisely, we want to study the chiral deformation as  the perturbation
of the
chiral symmetries including the conformal one. Because of the   lack of
a manifestly
chiral invariant action in general, we develop a new way of deforming
chiral symmetries.
 The novelty in this way of deforming is: first, it does not need to
first
know  the nature of the deformation field operator as we have the
habit   to do in
the conformal deformation; second it can be viewed as a constrained
method to deal with the deformation of general chiral symmetries.
 
The paper is organized as follows: In section 2 we recall some generalities
on
chiral algebras. The section 3 is devoted to discussing deformations of
chiral algebras and introducing our proposal approach. In section 4,
we apply our way of deforming the
 Virasoro algebra. We find a general class of isomorphism of
 this algebra. We show also that, starting from the highest weight
 representation with the vacuum state $|h>$, we find that the energy of
 this state is scaled by a real factor. The unitarity
constraints are worked out and as expected some known results are reproduced as special 
cases.
 
\section{Chiral symmetries.}
To start, suppose that we have a theory described by a hypothetical action
$S_0$ that
has a certain symmetry generated by a set of field operators
$\{ J_{\pm s},\theta_{\pm
(s-2)}, s=2,3,\dot \} $ satisfying the following conservation laws:
\begin{eqnarray}
& &\bar \partial J_s =\partial \theta_{s-2} \nonumber \\
& & \partial \bar J_{-s}=\bar\partial \theta_{-(s-2)}
\end{eqnarray}
this symmetry is said chiral if the field operators $\theta_{\pm (s-2)}$
are required
to vanish, that is :
\begin{eqnarray}
& &\bar \partial J_s =0 \nonumber \\
& & \partial \bar J_{-s}=0
\end{eqnarray}
Note that the constraint eq (4) are in fact integrability conditions
for the existence of the following Laurent development:
$$\quad \quad \quad \quad \quad \quad \quad \quad \quad \quad \quad \quad \quad \quad J_s (z)= \sum_{n\in Z}  z^{-n-s} J_n^{(s)}\quad \quad \quad \quad \quad \quad \quad \quad \quad \quad \quad \quad \quad  (4.b) $$
 
and a similar formula for the antichiral current $\bar J_{-s} (z)$.
Note also that the field operators $J_{s} (z)$ can be thought of as:
\begin{itemize}
\item Basic chiral current as in $W_N$ symmetries; $i.e$ $J_s=T(z)$,
$J_s=W_{(s)},\  ... $
\item Descendant field operators of primary ones.
\item  Composite operators. For the family of the identity operator
for example,, we may have in addition to
$J_s=T(z)$, the non linear composite operator:
$$J_s(z)=:T^2(z): $$
\end{itemize}
 
Before giving our way of deforming the chiral symmetries, let us  first
review briefly
how things are done in the conformal deformation. There, one start
from a critical
model described by an action $S_0$ and deform it to $s(\lambda)$ as
 
$$S(\lambda)=S_0 + \lambda\int d^2z \phi(z,\bar z)  $$
 
Where $\lambda$ is a perturbation parameter and $\phi$ is a given
relevant conformal
field operator. The original manifest conformal symmetry generated by
$T(z)$ and $\bar T (\bar z )$ is deformed to:
\begin{eqnarray}
& & \bar \partial T = \partial \theta \nonumber \\
& &  \partial \bar T =\bar \partial \theta
\end{eqnarray}
where the trace $\theta$ of the energy momentum tensor is proportional
to the perturbation $\lambda \phi(z,\bar z)$. Non manifest symmetries of
$S_0$ obeying
equation type (4) are in general deformed to equation of the form:
\begin{eqnarray}
& & \bar \partial J_s= B_{s-1}\nonumber \\
& &  \partial \bar J_{-s} =\bar B_{-(s-1)}
\end{eqnarray}
Different situations may happen here:
\begin{itemize}
\item[\bf{i}] Some of the $B_{s-1}$'s are still zero and
this means that some non manifest infinite subsymmetries  remain
unbroken. This is the case of the Crossover phenomenon.
\item[\bf{ii}] The
operator $B_{s-1}$ is a total derivative $B_{s-1}=\partial \theta_{s-2} $
and this means that we have at last one constant of motion $P_{s-1}$;
$$ P_{s-1}= \int dz T_s +\int d\bar z \theta_{s-2} $$
that remains conserved after deformation.
\item[\bf{iii}] The $B_{s-1}$'s are neither zero nor
total derivatives. In this case, the original non manifest symmetries of
$S_0$ are
completely destroyed by the deformation. A part from the Zamolodchikov
counting
argument which is based on analysis of vector spaces dimensions and
which gives some
indications on the conditions of occurrence of eq(3), there are no
systematic method
dealing with the three above mentioned situations. In case where
there are a large
number (an infinite number for the axiomatic point of view) of
$B_{s-1}$ satisfying
eq(3), we have an infinite number of constants of motion and the
deformation is said
integrable. This is what happens for example in magnetic and thermal
deformations of $c=1/2$ critical Ising model \cite{mussorado,fateevc}.
\end{itemize}
\section{Deformation of chiral symmetries.}
From the point of view of representation theory, the deformation of
the chiral
symmetry can be done in the same way as in the conformal case.  One has
to identify the
relevant field operators and the chiral symmetry we want to deform.
 Of course in general  we
  do not know how  how to write action exhibiting a manifestly chiral
invariance, but we can still study the deformation of any chiral
invariance, but we can still study the deformation of any chiral
symmetry just by manipulating the degree of freedom of the theory.
Among the numerous examples, one can quote the two following.
 Both these two examples concern the deformation of the $c=6/7$
critical theory. Besides the conformal invariance, the above mentioned
theory, known also as the tricritical  three states Potts model,
exhibits   two chiral symmetries: a W--symmetry whose current 
W has a conformal weight $\Delta=s=5 $ and a fractional
superconformal symmetry generated by the chiral current $G_{4/3}$  of
weight $\Delta=s=4/3$.
Using the fusion algebra of the $c=6/7$ critical model,   it can be
shown   that W-symmetry and the fractional superconformal  one lead
respectively to the following conservation laws: \begin{eqnarray}
& & \bar \partial W_5=\partial \theta_3 \nonumber\\
& &  \bar \partial G_{4/3}=\partial \theta^{'}_{-2/3} \nonumber
\end{eqnarray}
where $\theta_3=\lambda \phi_{22/7,1/7}$,
$\theta^{'}_{-2/3}=\lambda \phi_{1/2,15/21}$. In these equations,
$\phi_{h,\bar h}$ are Virasoro primary fields of conformal weight
$\Delta = h+\bar h$ and conformal spin $s=h-\bar h$. $\lambda$ is a
perturbation parameter carrying a certain mass scale.
 
From the complex analysis point of view, the deformation of the chiral
symmetry defined by eqs (4) can be achieved by deforming the Laurent
expansion eq(4b). Our motivation in doing so is that chiral symmetry
is the integrability principle for the existence
of Laurent development. Accordingly the deformation of chiral
symmetry is expected to induce a deformation of the Laurent expansion.
Let us write this deformed series as: \begin{equation}
J_s(z,\bar z) = \sum_{z\in Z} z^{-n-s} J_{n}^{(s)} (z,\bar z)
\end{equation}
where now the mode operators $J_{n}^{(s)} (z,\bar z$ carry some $z$
and $\bar z$ dependence. In perturbation   theory of parameter
$\lambda$,    these deformed Laurent modes may be written to the first
order in $\lambda$ as
\begin{equation}
 J_{n}^{(s)} (z,\bar z) =  J_{n}^{(s)} +{\cal O}_{n}^{s} (z,\bar z) +
 o(\lambda^2)
\end{equation}
where $J_n^{(s)}$ are as in eq(4.b) and ${\cal O}_{n}^{s} (z,\bar z)$
are some mode operators carrying the effect of the deformation.
 
Moreover using the fact that after deformation, eq(4) transforms in
general to
\begin{equation}
\bar \partial J_s(z,\bar z) =B_{s-1}(z,\bar z)
\end{equation}
Parameterizing the $B_{s-1}$ operator fields as a series:
\begin{equation}
B_{s-1}(z,\bar z) = \sum_{n \in Z} z^{-n-s+1} B_n^{(s-1)} (z,\bar z)
\end{equation}
then putting this expansion back in eq(9) and using the development of
$J_s(z,\bar)$, one gets: \begin{equation}
B_n^{(s-1)}(z,\bar z) = z^{-1} \bar \partial J_n^{(s)} (z,\bar z)
\end{equation}
In the case where $\bar \partial J_n^{(s)}\neq 0$ and $B_n^{(s-1)}=
\partial \theta_{n-1}^{(s-2)} $< the constraint eq(11) reads as:
\begin{equation}
\partial \theta_{n-1}^{(s-2)} =z^{-1} \bar \partial J_n^{(s)} (z,\bar z)
\end{equation}
The solving of eqs (11) and (12) can be brought to the problem of solving equation of the form: 

$$\quad  \quad \quad \quad \quad \quad \quad \quad \quad \quad \quad \quad \quad \quad \quad \quad \partial \Lambda =\bar
\partial \Gamma \quad \quad \quad \quad \quad \quad   \quad \quad \quad \quad \quad \quad \quad \quad  \quad \quad  (12.b)$$
where $\Lambda(z,\bar z)$ and $\Gamma(z,\bar z)$ are complex functions of $z$ and $\bar z$. We have looked for solutions of the above equation and we 
  have found that there exists  a class of solutions depending on a
 positive parameter $N$.
These solutions which read as
\begin{equation}
\Lambda =(\sum_{r=0}^{N} (-1)^r \frac{z^{r+1}}{(r+1)!} \partial^r )
\bar \partial \Gamma + \phi (\bar z )
\end{equation}
are elements of a set of functions $\Gamma$ defined by the constraint eq:
$$\quad \quad \quad \quad \quad \quad \quad \quad \quad \quad \quad \quad  \quad \partial^N \bar\partial \Gamma =0;  N=0,1,2....\quad \quad \quad \quad \quad \quad \quad \quad \quad \quad \quad \quad \quad (13.b) $$
where $\phi (\bar z )$ is an arbitrary antianalytic function 
($\partial \phi =0$).\\
Whether or not this constraint equation is related to the degenerate
equation \cite{polyakov} of the relevant conformal field operator of
the deformation in field theoretical approach remains an interesting
question that should be answered. Development in
this direction will be reported elsewhere.
Note that the field
operator    $\Gamma$ may expressed in term of $\Lambda$ by making the
change $z$ in $\bar z$.

Moreover, we note that eq (12.b) is in fact a special case of a more general one $$ \bar \partial U =A U + B \bar U + F $$
where $A,B$ and $F$ are continuous functions of $z$ and $\bar z$. 
The general solution of the above equation has been studied in the mathematical literature (
for details see \cite{comp1, comp2} ). In the case of our interest $A=B=0$, $U=\Gamma $ and $F=\partial \Lambda$, the integral solution reads as:

$$ \Lambda(z,\bar z) =\phi(\bar z) + 
\frac{1}{2i\pi}\int_D d^2 \omega 
\frac{\bar \partial \Gamma}{\omega - z } $$
where $D$ is the domain in which eq(12.b) is investigated and where 
$\phi(\bar z)$ is an arbitrary antianalytic function. Note that this form of the solution does not depend on any constraint on $\Gamma$ 
contrary to the series form solution eq (12.b). By considering constraints eq (13.b) the equation $\partial \Lambda =\bar \partial \Gamma $ has a solution in series form given by eq (13).
 
Note moreover that in  the case where the $J_n^{(s)}$ are $\bar z$
independent, the $B_n^{(s-1)}$'s operators are identically zero.  The
deformation eq(9) correspond simply to conformal transformations
preserving the chiral symmetry. In this case the deformed Laurent 
mode operators $J_n^{(s)}(z)$ may be written in terms of
the undeformed Laurent modes $J_n^{(s)}$ as:
\begin{equation}
J_n^{(s)} = f_n(z) J_n^{(s)}
\end{equation}
Expanding $f_n(z)$ in Laurent series and using eqs(7) together with eq(12),
one gets
\begin{equation}
J_s(z)=\sum_{n\in Z} z^{-n-s}  \sum_{m\in Z} f_{n,n-m}^{(s)}
j_m^{(s)}
\end{equation}
Choosing the Laurent modes $f_{n,r}$ as $f_{n,r}=\delta_{r,0} f_{n,r}$;
the above equation reduces to:
\begin{equation}
J_s =\sum_{n\in Z} z^{-n-s}  f_{n,0}^{(s)} j_m^{(s)};
\ \ \ \ f_{n,0}^{(s)}=f_n^{(s)}\neq 0, \ \ \ for\ all\  n.
\end{equation}
Two type of deformations should be distinguished here. As we shall
illustrate on the example of the Virasoro algebra, the deformations
obeying $f_{n+m}=f_n f_m$ describe automorphisms of the chiral
symmetry   and the deformations satisfying the condition $f
_{n+m}\neq f_n f_m$ give a general class of transformations   describing
isomorphisms of the chiral symmetry.
 
\section{Discussion and conclusion}
We have analyzed the case where the $J_n^{(s)}$'s are $\bar z$
independent at the level of Virasoro algebra, we have found that the
deformed algebra is given by the following parametric algebra:
\begin{equation}
\lbrack L_n , L_m \rbrack_f =\frac{f_{n+m}}{f_nf_m} (n-m) L_{n+m}
+c_e^f (n,m)
\end{equation}
where $c_e^f(n,m)$ is the central extension which depend on the non
vanishing complex function $f$. It is not difficult to show that the
bracket $\lbrack L_n , L_m \rbrack_f$ satisfy the antisymmetry relation
and Jacobi identity.
 
Now let us discuss the central extension $c_e^f(n,m)$ of this algebra.
To determine $c_e^f(n,m)$  we follow the method of the Goddard et al
\cite{goddard}. The central extension corresponds to a shift of the
vector space and therefore the central extension must not break the
antisymmetry of the bracket and Jacobi Identity. Using this fact, we
show that the central extension depend explicitly on $f$ and take the
following form:
\begin{equation}
c_e^f(n,m)=\frac{c}{12}\frac{n (n^2-1)}{f_nf{-n}}
\delta_{n+m,0} \end{equation}
When imposing unitarity, the function $f$ is constrained to:
\begin{equation}
f_n^* = f_{-n}, \ \ \ \ for \ all \ n\in Z
\end{equation}
where $f_n^*$ mean the conjugate of $f_n$.
It's obvious to see from this constraint that $f_0$ is real and if $f_n$
is a real function then it's necessary an even function (i.e
$f_{-n}=f_n$ ).
 
To study the highest weight representations of this deformed algebra, we
consider a
highest weight representation with the vacuum state $|h>$. It satisfies:
\begin{equation}
L_0|h>=h|h>, \ \ \ \  L_n|h>=0, \ \ \ \ n>0.
\end{equation}
We follow the method described by \cite{goddard},
from the eqs (17,18), we evaluate first $<h|L_{-n}^+L_{-n}|h>$:
and found that it's given by:
\begin{eqnarray}
<h|L_{-n}^+L_{-n}|h>&=&
    \frac{1}{|f_n|^2} ( 2 n h f_0+ \frac{c}{12}n(n^2-1) ) <h|h>
\end{eqnarray}
The quantity $<h|L_{-n}^+L_{-n}|h>$ must be positive if the Hilbert
space has a positive norm. One conclude then that
$2 n h f(0)+ \frac{c}{12}n(n^2-1) >0$ for all $n>0$.\\
From the last equation and for $n=1$ one deduce that $hf(0)>0$. For
large value of $n$ this constraint give $c>0$. Such representation is
then caracterised by $hf(0)>0$\footnote{Rather than the condition $h>0$
for Virasoro algebra, in this case we have to re-scale the parameter $h$
by a factor $f_0$} and $c>0$.
We have also shown that in the case of real function $f_n$, the Verma
determinant of $N=2$ and $N=3$ representations take the same form as in
the Virasoro algebra with changing $h$ with $h f_0$.
 
In the case where $f_{n+m}=f_n f_m$ one see from this condition that
$f_0=1$. With these two constraints one reproduce exactly the known
result on the automorphisms of the conformal algebra \cite{patera}.

\end{document}